# Direct current conditioning to reduce the electrical impedance of the electrode to skin contact in physiological recording and stimulation

Pavel Govyadinov [1, 2], Sergei Turovets[2], Amanda Gunn[2], Don Tucker[2], Phan Luu[2]

[1]Department of Computer Science, University of Oregon

[2]Electrical Geodesic Inc., Eugene, Oregon.

**Abstract**

**Direct and alternating current iontophoresis and electro-osmosis methodologies have provided new methods of transcutaneous drug delivery. A byproduct of such methods is lowering the electrical impedance of the electrode to skin contact, as conductive ions permeate the stratum corneum, the primary resistive layer of the skin. We developed a method for adapting iontophoresis to condition the electrode to skin contact, both for electrophysiological recording and electrical stimulation of body tissues. By utilizing direct current to treat electrodes with high impedance we show the effectiveness of iontopheresis as a driving force for permeation of ionic electrolyte into the skin barrier. We applied direct current (DC) levels of 50 µA to electrodes on the human head for 30 seconds with paste (Nihon Kohden Elefix) electrolyte. Typically immediately after DC treatment conditioning there was an impedance drop of 10-30%. The effect was lasting over several hours, with the paste electrolyte. These results demonstrate the feasibility of DC conditioning to reduce the set time of electrolytic solutions and to maintain good skin contact during extended recording or stimulation sessions.**

1. Introduction.

   In the early methods of electrophysiological recording, such as with electroencephalography (EEG), low electrode-scalp impedance was necessary, primarily because of the relatively low input impedance of the tube amplifiers. In addition to using electrolytes to hydrate the skin, it was common to abrade the skin (scalp) to reduce impedance. Breaking the skin barrier (the most resistive and protective layer, the stratum corneum) is painful, and it increases the risk of blood-borne infections, such as hepatitis C or HIV. A common early misconception was that higher electrode-skin impedance would lead to attenuation of the EEG or other physiological signal. With modern solid-state instrumentation amplifiers, input impedances of hundreds of Megohms (or even > 1 Gigohm) are common, allowing better tolerance of high electrode-skin impedance (Ferree et al., 2001). Since amplitude attenuation is a function of the ratio of the electrode-skin impedance to the amplifier input impedance (Ferree, et al., 2001) with modern recording methods, even an electrode-skin impedance of 100 KOhms remains a small fraction of a modern amplifier input impedance often exceeding a hundred megohms. Finally, the typical environmental noise induced by high electrode-skin impedance mismatch in channels of a differential amplifier is caused by electromagnetic interference of local power circuits, such that it is concentrated at a narrow frequency (50 Hz or 60 Hz) and is easily filtered from the physiological recording.

   Nonetheless, physiological recording with the low amplitude (microvolt) signals, such as EEG, remains challenging with high impedance contacts, and improvements are desired. If saline-sponge electrolytes are used, drying of the saline in sponges and around an electrode shrinks the area of contact and leads to unacceptable (> 200 K Ohms) impedances within one to two hours, depending on humidity. If a paste electrolyte, such as Nihon Khoden's Elefix, is used, extended recordings are possible due to a stable contact area. However, the initial passive hydration of the skin is slow, such that electrode-skin impedance declines slowly over an hour or more reaching a minimum and only afterwards a process of drying takes over leading to a gradual impedance increase.

The high magnitude, mismatch and temporal variations of electrode to skin impedance can also cause loss of common mode rejection and problems for signal integrity and degrade data quality in Electrocardiography (ECG) (Huhta and Webster 1973, Medina et al 1989, Oster 2000, Miller et al 1985), Electromyography (Zipp 1982) and Electrical Impedance Tomography (EIT) (McAdams et al 1996). The high contact impedance can lead also to impressed current amplitudes clipping with the limited voltage budget and failure to deliver to the target the needed amount of current density in transcranial direct current stimulation (tDCS) (Hahn et al 2013 ).

The noninvasive ways of improving permeability of the stratum corneum and eventually improving the electrical contact with the skin are well known in the related field of the transdermal drug delivery enhancement (Fox et al, 2011). Different types of skin penetration enhancement techniques involving chemical (Fox et al 2011) and physical processes (like temperature, ultrasound or direct current) have shown the potential to reversibly overcome this barrier to provide effective delivery of drugs across the skin. However, the chemical enhancers may be toxic or skin irritating and require further study before the wide introduction into the clinical practice. On the other hand, control of skin-electrode interface through temperature or applying ultrasound in many circumstances are not practical. Iontophoresis, the physical process of driving ions through a medium with electrical current (Prausnitz 1996), seems to be the most promising technique for electrical contact conditioning. At the present time, the primary application of iontophoresis is drug delivery through the skin (Kalia et al, 2004) and electrode to skin impedance drop during ionthoporesis is used primarily as an auxiliary metric for skin permeability enhancement. The application of an electric current to skin tissue reversibly alters the barrier properties of the well hydrated skin (Dhote et al, 2011). The positively charged ions in an ionic substance are repelled away from the anode, while the negatively charged ions are repelled from the cathode and driven across the skin entraining the electrolyte through electro osmosis and facilitating skin conductivity. The improved concentration of electrolyte in skin tissue increases the electrical conductance (decreases electrical impedance). Oh and Guy (1995), for example, showed that application of electrical current after 2 hours of passive hydration decreases electrode-skin impedance reversibly from 10% to 45% from the initial level, depending on the amplitude of the current and the length of treatment. The largest impedance change is observed in the first 60 seconds of the iontophoretic application, with the majority happening in the first 10 seconds (Oh and Guy, 1995). When skin is initially dry, electro osmosis (Grimnes 1983) usually dominates over iontophoresis and an initial resistance drop is usually irreversible. In real applications, the resistance drop is most likely due to the mixture of both effects.

In the present study, we developed an initial protocol for intentional iontophoretic/electro osmotic conditioning of the electrode-skin contact. The goal was to decrease the contact impedance to improve physiological signal quality and reduce noise, in this case for EEG. We applied one current level of 50 µA (which was imperceptible in our case of approximately 100 $mm^2$ EEG electrodes) for 30 seconds. Paste electrolyte was examined. With the goal of a practical routine protocol, we examined the initial impedance drop within a few minutes after current application (suitable for a practical implementation after a brief conditioning period during the initial set up of an experiment) as well as the stability of the impedance change over several hours and response to repeated DC conditioning in the end of a long session. Two minutes of eyes-closed resting task was performed at the beginning and at the end of the experiment in order to show a reduction in noise over time due to a reduction in average electrode to skin impedances.

2. **Materials and Methods.**
   *2.1 Experimental Setup.*

   We employed two platforms based on the multichannel NA 300 EEG amplifier and NA 400 EEG amplifier with a built-in DC/AC current generator and Net Station v4.4.1a1 or Net Station 5.1.1 software (Electrical Geodesics, Inc.) to perform DC conditioning and AC electrode to skin impedance measurements. The subjects were wearing a 256 electrode Geodesic Sensor Net (Electrical Geodesics, Inc.) with a paste electrolyte (described below) and sitting in a comfortable position in a chair (Fig. 1). The NetAmps 300/400 platform synchronously digitizes 256 analog channels at 20 kHz and 24 bits, and uses a field-programmable gate array (FPGA) to collate and transfer data in IEEE 1394 (Firewire) format to a computer. The NetAmps 300/400 platform uses "sigma-delta" type A-to-D converters, achieving high linearity, accuracy and resolution at low cost, so that each sensor can have a dedicated A-to-D converter, without the need for high-speed multiplexing of different electrode signals onto a single higher-speed A-to-D converter. The current source is based on a Howland-type design and isolated from the amplifier circuitry and was used earlier in the similar experimental setup in the context of low frequency EIT (Esler et al., 2010). Advantages of the Howland topology over other designs include its small number of components, a single active device and the ability to adjust output resistance. The waveform and level of the injected current is monitored across a 100 Ohm resistor in series with the impedance load containing the subject head and injector/sink electrodes. A scanning protocol, including specific sequence of source/sink electrodes, frequency (DC or AC in the EEG range, typically 1-400 Hz in the AC mode), current amplitudes, and duration of epochs is programmable via a graphic user interface and allows to swipe any of these parameters. The impressed potentials are recorded as regular EEG signals and their amplitude and phase are extracted through a software implementation of locked-in detection. The bandwidth of the current injector is set from 0.5 Hz to 500 Hz by the low pass hardware and digital filters in the NetAmp 300/400.

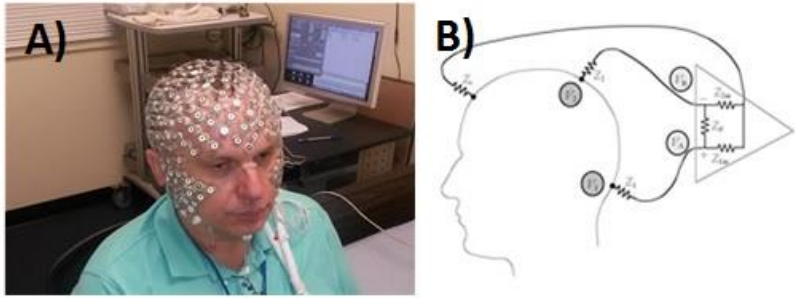

*Figure 1: Experimental setup: (left) EGI's 256 Geodesic Net on subject's head connected to the NA 300, (right) Schematic of DC/AC voltage impedance measurements with a differential amplifier NA 400.*

   *2.2 Impedance Measurements.*

   Since making an independent measurement of the potential just below the scalp surface is impractical, an approximation method is required. We used the standard impedance checking methods in EGI NA 300/400 system (shown in Fig. 1 from Ferree et al 2001). Impedances were measured from each electrode in the net using EGI NA300/400 amplifier and Net Station v4.4.1a1 or Net Station 5.1.1 software. By driving all but one (to be measured) of the electrodes to a known potential relative to the ground (rms 400 μV at 20 Hz) and estimating the current flowing through the resulting circuit, one can estimate the electrode impedance as all other channels are connected in parallel and their total resistance is 255 times less than an average 50 kOhm electrode-to-skin resistance. The total head tissue resistance is estimated to be below 0.5

kOhm and also can be neglected. The remaining circuit including a 10 kOhm resistor of the measuring channel to the ground is a simple voltage divider, so the unknown impedance, Z, is given by Eq. (1).

$$Z = \frac{10.0\ kOhm * 400\mu V}{V_{osc}} - 10 kOhm$$

(1)

Here $V_{osc}$ is the oscillatory signal rms amplitude between the electrode to be measured and ground determined by software means from several digitized signal wavelengths. This process is fast and takes approximately 5 seconds to measure impedances of all 256 signal and one reference electrodes to skin in the standard 256 Geodesic Sensor Net.

*2.3 Nihon Kohden Elefix Paste.*

Elefix (Nihon Kohden America, Inc. 15353 Barranca Parkway, Irvine, CA 92618) was used in all human trials in this study. It is a propylene glycol based paste electrolyte that is FDA cleared and available commercially.

*2.4 Current Impression Protocol.*

Eleven subjects (4 females and 7 males) with the ethnic background corresponding to the Eugene-Springfield, OR metropolitan area, having a wide range of the skin and hair types, and aging from 19 to 70 years old were recruited for the study of iontophoretic conditioning of EEG electrode to skin contacts. The protocol was approved by Electrical Geodesics, Inc. Institutional Review Board (IRB) as safe and ethically appropriate for human subject tests.

Two different protocols were used: a protocol to study the immediate impedance drop due to iontophoretic treatment (6 males and 3 females) and a long term current injection protocol to show the longevity effect of the conditioning (one male, one female). A dry 256-channel Geodesic sponge-less net was applied and each electrode pedestal was then filled with Elefix paste from a tube warmed in hot water.

In the first protocol special care was taken in order to ensure that there was no excess electrolyte in order to minimize the area of the current injection. After application of the EEG electrodes, initial impedance values were measured. This was followed by injection of DC current between 12-16 source-sink pairs per run with one electrode acting as a sink (cathode) and one electrode acting as a source (anode). Source-sink pairs were chosen such that their physical location was approximately opposite on the head in relation to one another. DC current was then injected through the source-sink pairs at an amplitude of 50µA for 30 seconds. Following each 30-second injection, impedance data were again collected using the same methods. Current was never applied to the same source sink pair twice during the same experiment. The remaining electrodes on the 256-channel Net were not conditioned with current, but their impedances were measured to provide a baseline of untreated electrode-skin contacts.

In the second protocol application followed the standard longevity procedure. A two minute resting task was added in the beginning and end of the session and the EEG was recorded at 1000 samples per second. After application and the first resting task, initial impedance values were measured. This was followed by injection of DC current between 16 source-sink pairs per run with one electrode acting as a sink (cathode) and one electrode acting as a source (anode). Source-sink pairs were chosen such that their physical location was approximately opposite on the head in relation to one another. DC current was then injected through source-sink pairs at an amplitude of 50µA for 30 seconds. Following each 30-second injection, impedance data was again collected using the same method. After the first current injection block, current was injected again using AC current (at 1Hz and 50 µA for 60 seconds) through a separate set of 16 source-sink pairs. Finally, DC current was injected again using 50 µA for 30 seconds through yet another set of 16 source-sink pairs. Impedance data were again collected after each individual source-sink pair conditioning. At the end of the experiment the subject performed a two minute resting task. The complete experiment lasted for 156 minutes.

Nine experiments were performed in total using the first protocol, and two experiments were performed using the second protocol.

## 3. Results.

Data analysis focused on three aspects of DC iontophoretic conditioning: 1) the initial impedance drop for source and sink electrodes, 2) the longevity of the effect (stability over time) and 3) the effect of lowered impedance on the frequency spectrum during the EEG resting task. We only considered the longevity of the effect in cases where an initial effect was observed.

*3.1 Longevity.*

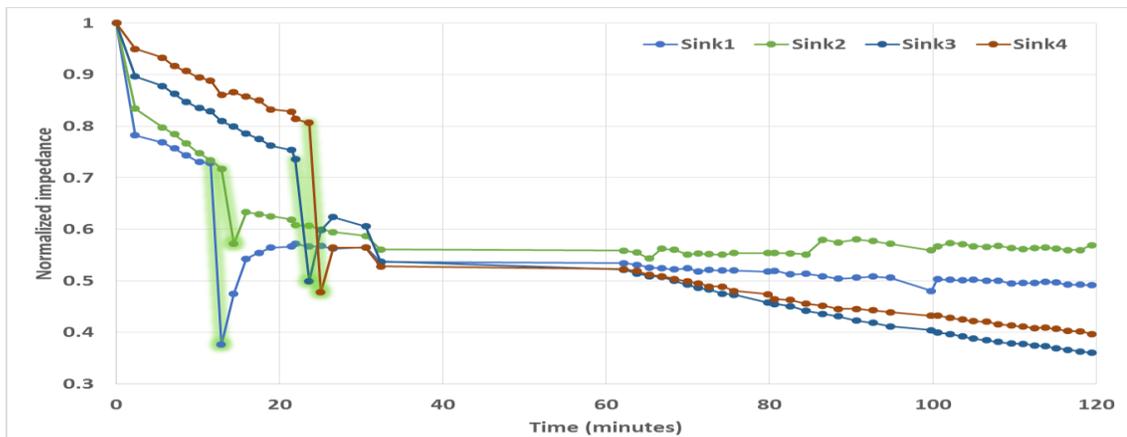

*Figure 2: Impedances over time normalized to impedance at t=0 minutes for an assortment of sinks from subject 1 long term session. Highlighted regions show the period during which the treatment happened. There was a 30 minute break in the session between 32 and 63 minutes to provide respite for the subject.*

In the two long-term sessions, each with a different subject, we considered the longevity of the electrode conditioning effect. As a method of visualization over time, we produced two figures displaying the impedance of a subset of DC treated electrodes as compared to the mean impedance. The Elefix paste experiment showed a lasting iontophoretic conditioning effect, with the impedance drop lasting for the duration of the entire experiment in most cases (Fig 2, 3). All impedance are reported in terms of impedances normalized to starting impedance of four sink electrodes from subject 1 (Fig 2) and their corresponding sources (Fig 3), with line colors designating source-sink pairs. Each electrode is treated once between t = 10 minutes and t = 30 minutes, the exact time point is highlighted. Both figures clearly show an impedance drop ranging from 10-30% from starting impedance due to a single treatment commonly with actual drop ranging from aproximately 60 kOhms (30% of impedance at t = 0) to 5 kOhms (10%). In all sink electrodes the large intial drop is followed by a relaxation period where the impedance returns to a higher value, further demonstrating the presense of both, electro-osmotic (irreversible) and iontophoretic (reversible) effects. The sources commonly showed similar behavior. In both cases the impedances continued to gradually decrease as the session continued due to passive hydration. Similar results were seen with subject 2.

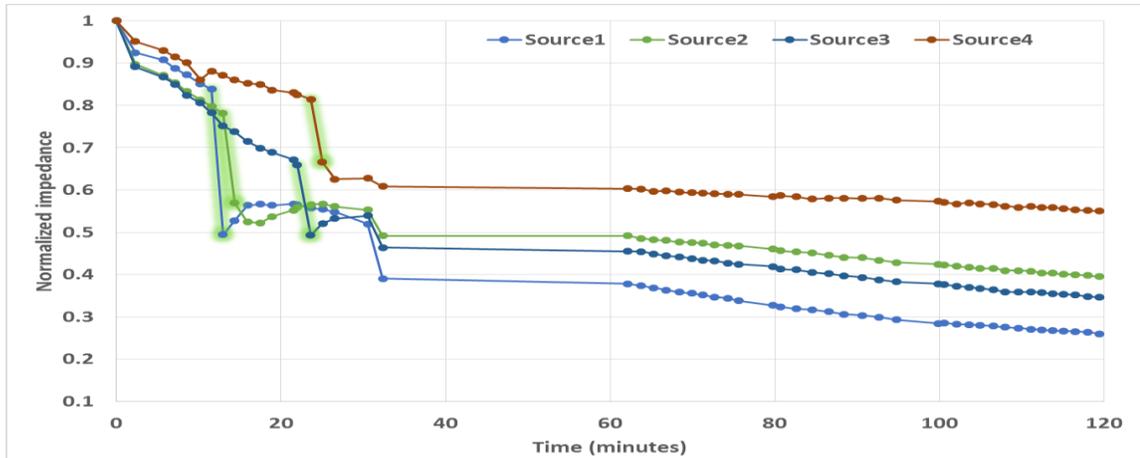

*Figure 3: Impedances over time normalized to impedance at t=0 minutes for an assortment of sources corresponding to the sinks from subject 1 session. Highlighted regions show the period during which the treatment happened. There was a 30 minute break in the session between 32 and 63 minutes to provide respite for the subject. Sources are color coded to sinks in Figure 2.*

### 3.2 Initial Effect in Short Sessions.

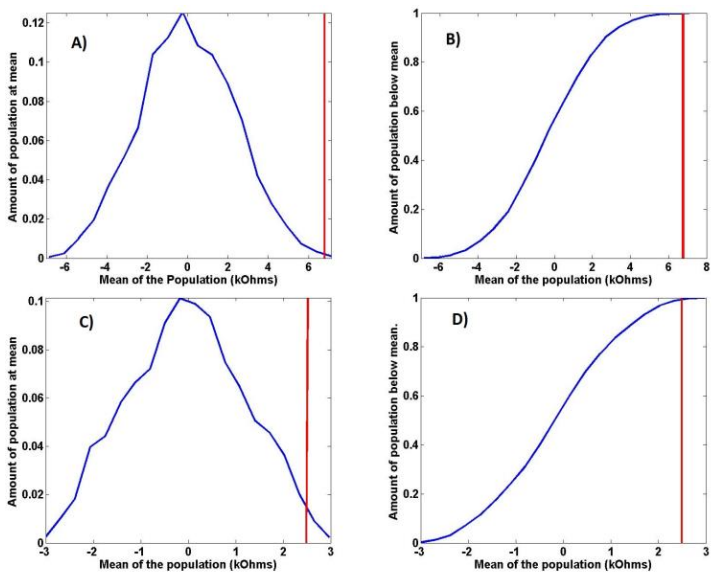

*Figure 4: Comparison of the random permutation samples, line in red represents the location of the observed value. Sources and sinks show statistical significant at 0.05 level. a) Distribution created by resampling the sink and untreated sets, b) The observed data falls in the top 99% of the randomly created population using the sink set, c) Distribution created by resampling the source and untreated sets d) the observed data falls in the top 96% of the randomly created population using the source set.*

Due to the relatively low sample size (9) and the large discrepancy between the size of the treated population (16 sources and 16 sinks) and the untreated populations (≈224 electrodes) a non-parametric permutation test based on the mean decrease of either the sink or the source channels was chosen to be the best way to demonstrate statistical significance. Each data set for the nine subjects was analyzed in respect to three values: a mean impedance change across all the source electrodes, a mean change across all sink electrodes, and a mean impedance change across all untreated electrodes. Two subsets were created out of the resulting data, one comparing the sources versus the untreated group, one comparing sinks versus the untreated group. The 5000 random permutation sample sets were generated, creating a quasi-Gaussian distribution, which were then compared with the observed values.

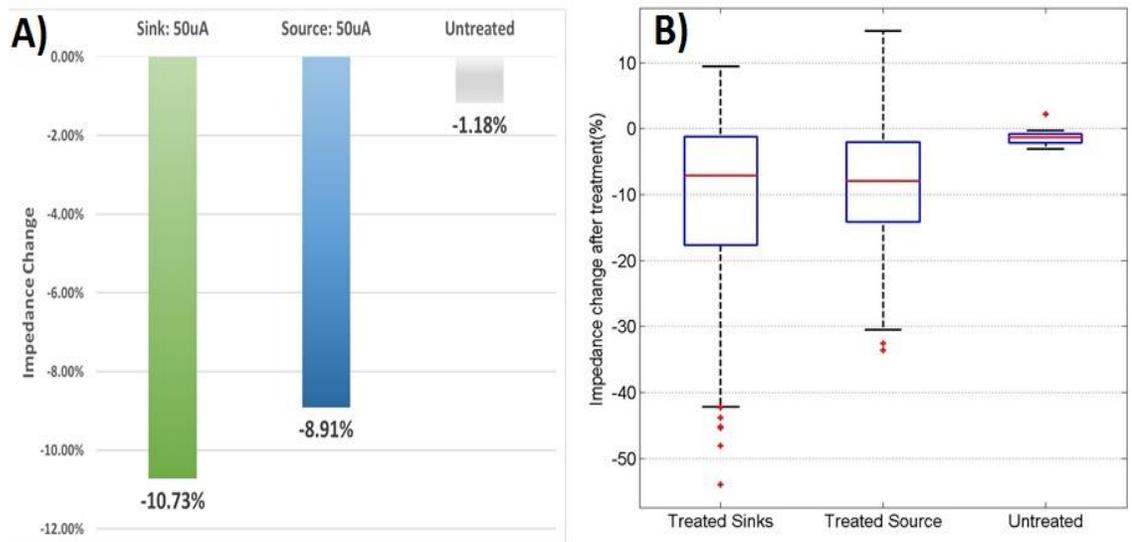

*Figure 5: Relevant statistical information on 9 treated subjects. a) The mean impedance change across 288 treated source or sinks electrodes and ≈2016 untreated electrodes, b) Box and whisker plot has the distribution information, the red line signifying the median value.*

As a result of this analysis, we found that both the source and sink sets were in the top 95% of the random permutation samples, meaning that the treatment of the source and the sinks showed a statistically significant impedance drop over untreated electrodes with the significant level of 0.05. (Fig. 4). We also found that an average of 10% drop can be observed throughout the data, with the maximum decrease of 53% (Fig. 5).

*3.3 Power Spectral Analysis.*

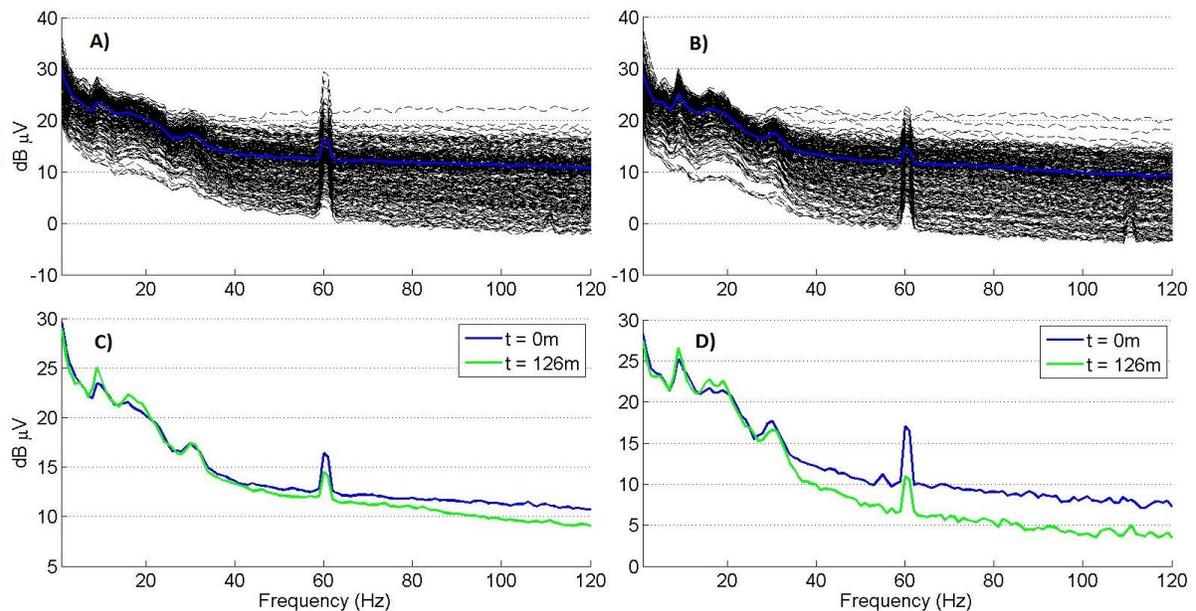

*Figure 6: Power spectrum density analysis. a) Frequency spectrum of the resting EEG (160 electrodes) prior to any sort of treatment with the electrodes passively hydrating the skin for approximately 30 minutes (t = 0, mean impedance = 76kOhms), b) The frequency spectrum of the EEG resting task (160 electrodes) with electrodes passively hydrating the skin for approximately 156 minutes (t = 156, mean impedance = 30 kOhms), c) Comparison of the mean PSD for 160 electrodes at t = 0 and t = 126 minutes, d) Comparison of the PSD of a typical channel, at t = 0 and t = 126 minutes.*

Two minutes resting EEG time series were quantitatively processed with the standard spectral power approach. The MATLAB pwelch function was used to process the two minute time series which was subdivided into 1000 ms epochs in order to provide the required frequency resolution. The resulting spectra were averaged over all the epochs. Figure 6 shows the spectra of individual channels (black) and the grand average over all the channels (colored), as well as showing the difference between the grand averages of the two minute resting EEG time series at the beginning of the session and the end of the session (Fig 6, bottom). Only frequencies from 1 to 120 Hz were analyzed. Prior to analysis a band-pass filter (1 to 120 Hz) using NetStation 5.1.1 filtering software was used to clean each time series. A 60 Hz notch filter was not applied in order to better demonstrate the improvement in frequency response due to falling impedances. By the end of the session the high/low-gamma background, i.e. 60-120 Hz and 60 Hz are lower (1.41 µV), likely due to lower impedances and reduced environmental noise-coupling. Note that all results reported in figure 6 are reported as $\log_{10}(\mu V^2/Hz)$ to better illustrate the effects of reduced impedance across the entire spectra.

4. Discussion.

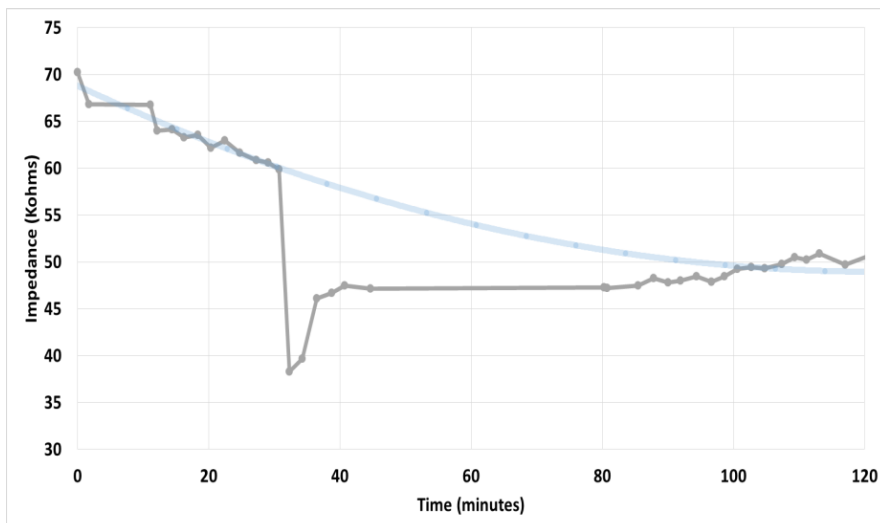

Figure 7: Impedance over time with a hypothetical polynomial trend-line demonstrating the possibility for quick hydration time. Data reported from a long term session for subject 1.

The present results suggest that iontophoretic conditioning can be applied to improve electrode-skin contact for physiological recording. The propylene glycol based Elefix provided adequate amount of ions for the process with EGI Geodesic Sensor Net electrodes. The currents were impressed over a safe, and comfortable range. 50 µA amplitude was large enough to cause a significant current drop, and the current was only slightly perceptible; shown only as a slight itching feeling rated 2 on average across all the participants (slightly uncomfortable but not painful).

The reasonable impedance decreases (from 1% to 53% for sinks electrodes and from 8% to 33% for sources) due to DC conditioning was observed with the Elefix electrolyte. Generally, the propylene glycol based paste (Elefix) provided stable impedances over time lasting over 2 hours in some cases even without DC conditioning. However this passive skin hydration process is slow (about 1 hour or more) and in addition to a tedious application procedure for the dense array of electrodes taking between 30 to 60 minutes, the total preparation time for an EEG session becomes unacceptably long. Our results show, that the DC conditioning can be particularly useful in improving the electrode-skin impedance quickly and can become a standard procedure in physiological recordings. In Fig. 7 we plotted a typical sink electrode impedance versus time and added an asymptotic trend line of progressing impedance with time in case of passive hydration. It demonstrates further how treatment can quickly reduce electrode-skin impedance. Additionally Figs 2,3, and 8 all show impedances continuing to gradually reduce post-treatment, suggesting that the tissue under the electrodes has not been saturated completely. This suggests that treating electrodes multiple times and/or increasing the length of the treatment may improve the results, pointing to further research.

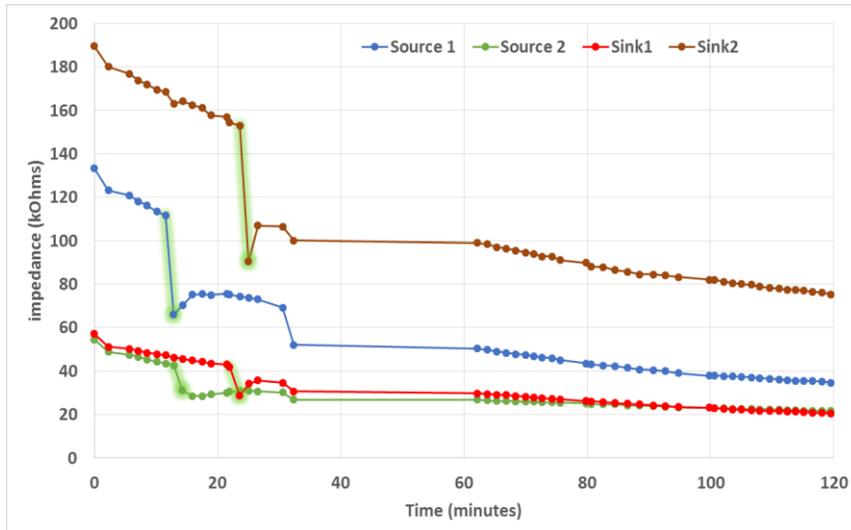

*Figure 8: A comparison of the effect due to DC treatment on electrodes with high absolute starting impedance versus low absolute starting impedance in kOhms.*

In this study, a consistently greater impedance drop was observed for the sink (cathode) electrode than the source (anode) electrode. One possible explanation may be rooted in contributions of electro-osmosis, which is greater at cathodal treatment (Grimnes, 1983). One can conclude that the positive ion entrainment of neutral molecules and electrolyte transfer from the anode through the stratum corneum (including sweat ducts, pores and newly established smaller pathways) and its resulting hydration was apparent ly lower than transfer of electrolyte associated with negative ions repulsion through the skin at the anodal electrode and skin hydration from the cathodal site. Yet another explanation can be in a skin physiological response to the negative ions at the cathodal side, for instance sweat ducts expansion. For DC conditioning, it may therefore prove useful to switch the anode and cathode during the procedure. Alternatively, each electrode may be used as a sink with multiple source electrodes, with this pattern varied systematically over the head. In addition, to further optimize the procedure, our research suggests that electrodes with high starting impedance (> 40 kOhms) tend to benefit more than electrodes with low starting impedance (Fig 8). Note that we see a very large impedance drop with electrodes that have yet to reach optimum impedance, while electrodes with starting impedance ≈40 kOhms do not retain their lowered impedance value. This is likely due to a difference in electrodes' physical location, i.e glandular density, hair density and pore density.

We also show that a moderate decrease in electrode-skin impedance shows decreased amplitude in the Low/High Gamma frequency response with a typical EEG signal during a resting task with eyes closed characterized by the peak in the alpha band (9-12 Hz). There is no clear effect in the delta (1-4 Hz), theta (5-8 Hz), alpha and beta (13-30 Hz) frequency bands, but a notable and consistent difference between the frequency amplitudes in the 40-120 Hz frequency range. Although this difference is slight in the data we collected, with the hardware being the most likely cause, since the amplifier used is specifically designed to decrease the sensitivity to electrode-skin impedance (Ferree et al, 2001), we still show an improvement in frequency response. DC treatment may be of even more use for physiological recordings where tissue abrasion is not an option and other methods must be used to further increase the signal-to-noise ratio (SNR), a necessity in applications where the signal is potentially weak, such as EIT.

Although good results were obtained with DC currents, the iontophoresis literature suggests that AC protocols may achieve similar impedance drops. In our initial AC experiments (Protocol 2) we have observed some impedance change with conditioned electrodes in 2 subjects at 50 uA and 1 Hz. In the literature, high frequency (10 KHz) AC at 50 µA has been shown to be effective in lowering impedance in studies with nude mice (Bagniefski and Burnette, 1990). AC frequencies in this range could be applied coincident with EEG recording, given that they are well outside the band-pass of the EEG filters. Furthermore, AC currents do not run the risk of polarizing the electrodes (although we observed minimal polarization with DC currents with the Ag/AgCl coated carbon-fiber plastic EEG electrodes used in this study). Our research points to electroosmosis as a significant driving force along with iontophoresis to have impact on electrode-skin impedance for physiological measurements (Grimnes, 1983) especially in the beginning of measurement when skin is relatively dry. Finally, further research and development that integrates digital signal analysis with the iontophoretic conditioning is needed to provide practical guidance of the protocol as a function of repeated measurement of the electrode-skin impedance, as well as protocol of treatment on individual electrodes, such as electrode location and optimal DC current amplitude and duration. Such feedback guidance and analysis will be particularly important for managing and accessing electrode-skin impedance in transcranial electrical stimulation (TES) protocols, as has been recently suggested by Hahn et al 2013, either with DC (tDCS) or AC (tACS) currents. The electrode-skin impedance must be assessed in planning the desired current impress pattern for the neurostimulation protocol.

## 5. Acknowledgements.

We thank Brian Esler who assisted in the preliminary work for this study. We are also thankful to Benson Kuo and Jasmine Song for discussion of processing EEG spectral data and non-parametric statistical tests. We point out that Electrical Geodesics, Inc., has applied for a US patent for the method of improving electrode to scalp electrical impedance for electrophysiological measurement or stimulation with iontophoretic conditioning.

## 6. References.